\documentclass[review]{elsarticle}

\usepackage{lineno,hyperref,amsmath,bm}
\biboptions{super,sort&compress}

\journal{arXiv.org}

\begin{document}

\begin{frontmatter}

\title{Velocity-gauge real-time TDDFT within a numerical atomic orbital basis set}

\author[tim]{C. D. Pemmaraju\corref{ca}}
\cortext[ca]{Corresponding author}
\ead{dasc@slac.stanford.edu}
\author[uwp]{F. D. Vila}
\author[uwp]{J. J. Kas}
\author[mpi]{S. A. Sato}
\author[uwp]{J. J. Rehr}
\author[uot2]{K. Yabana}
\author[tmf]{David Prendergast}

\address[tim]{Theory Institute for Materials and Energy Spectroscopies, SLAC National Accelerator Laboratory, Menlo Park, CA 94025, USA}
\address[uwp]{Department of Physics, University of Washington, Seattle, WA 98195, USA}
\address[mpi]{Max Planck Institute for the Structure and Dynamics of Matter, Luruper Chaussee 149, 22761 Hamburg, Germany}
\address[uot2]{Center for Computational Sciences, University of Tsukuba, Tsukuba 305-8577, Japan}
\address[tmf]{The Molecular Foundry, Lawrence Berkeley National Laboratory, Berkeley, CA 94720, USA}





\begin{abstract}
The interaction of laser fields with solid-state systems can be modeled efficiently within the velocity-gauge formalism of real-time time dependent density functional theory (RT-TDDFT). In this article, we discuss the implementation of the velocity-gauge RT-TDDFT equations for electron dynamics within a linear combination of atomic orbitals (LCAO) basis set framework. Numerical results obtained from our LCAO implementation, for the electronic response of periodic systems to both weak and intense laser fields, are compared to those obtained from established real-space grid and Full-Potential Linearized Augumented Planewave approaches. Potential applications of the LCAO based scheme in the context of extreme ultra-violet and soft X-ray spectroscopies involving core-electronic excitations are discussed.
\end{abstract}


\end{frontmatter}


\section{Introduction}
Over the last two decades, real-time time dependent density functional theory\cite{Runge1984,Marques2012a,Ullrich2012} (RT-TDDFT) approaches, wherein the electron density is explicitly propagated in time through numerical integration of the time-dependent Kohn-Sham equations, have gained in prominence as practical first-principles methods for studying electron dynamics in a wide range of quantum systems\cite{Marques2012a,Ullrich2012,Bertsch2000,Tsolakidis2002,Takimoto2007,Meng2008,Lopata2011,Castro2012,Andrade2012, Yabana2012,Wang2013,Krieger2015,Goings2016,Nguyen2016,Provorse2016,Yost2017}. By directly simulating density fluctuations in time\cite{Marques2012a,Ullrich2012}, RT-TDDFT provides a versatile and computationally tractable framework for accessing linear and non-linear response properties of materials as well as electron dynamics in conditions beyond the perturbative regime, for instance, under the action of intense ultrafast laser pulses\cite{Sato2015,Floss2017}. While a  variety of utilizations of RT-TDDFT have been demonstrated in recent years\cite{Ullrich2012,Bertsch2000,Tsolakidis2002,Takimoto2007,Meng2008,Lopata2011,Castro2012,Andrade2012, Yabana2012,Wang2013,Krieger2015,Goings2016,Nguyen2016,Provorse2016,Yost2017,Sato2015,Tancogne-Dejean2017,Floss2017,DeGiovannini2013,Sommer2016,Lucchini2016,Miyamoto2017,Hubener2017}, one of the most relevant application domains for RT-TDDFT is the study of laser-matter interactions thanks to the broad appeal of ultrafast laser spectroscopies as experimental tools for investigating and controlling excited states of matter\cite{Castro2012,Krieger2015,Yabana2012,Sato2015,Tancogne-Dejean2017,Floss2017,DeGiovannini2013,Sommer2016,Lucchini2016,Miyamoto2017,Hubener2017}. This is especially the case in a solid-state or condensed-matter context where RT-TDDFT represents perhaps the only computationally feasible first-principles approach for treating the action of intense laser fields\cite{Yabana2012}. Condensed phase implementations of RT-TDDFT\cite{Yabana2012,Krieger2015,Andrade2015} that rely on the periodic supercell framework treat light-matter interaction through the so-called velocity-gauge form $\mathrm{\overrightarrow{A}.\overrightarrow{p}}$ involving the vector potential and momentum operators, in contrast to traditional implementations for isolated atomic and molecular systems\cite{Tsolakidis2002,Takimoto2007,Lopata2011} where the length-gauge form $\mathrm{\overrightarrow{r}.\overrightarrow{E}}$ coupling the position and electric field operators is employed. Numerical implementations of velocity-gauge RT-TDDFT (VG-RT-TDDFT) based on real-space grids\cite{{A.Sato2014},Andrade2015} and full potential linearized augmented planewaves (FP-LAPW)\cite{Dewhurst2004} have been demonstrated in recent years and used to investigate laser-induced valence electron- and spin-dynamics\cite{Yabana2012,Krieger2015} and related phenomena in solid-state systems with encouraging results. 

Valence electron excitations which involve energy scales of a few eV are amenable to a very efficient treatment using uniform real-space based RT-TDDFT\cite{A.Sato2014,Andrade2015} as the valence electron density and its fluctuations are smoothly-varying in space and can be well represented on relatively coarse grids with grid spacings on the order of 0.5 a.u. Analogously, in planewave implementations\cite{Yost2017} a small kinetic energy cutoff on the order of a few tens of Rydberg is sufficient. In practical simulations that are primarily concerned with valence electron dynamics, one often employs either a pseudopotential or frozen-core approximation whereby the localized inner-shell core electrons are effectively eliminated from the description facilitating low-cost real-space or planewave expansions. However, if one is interested in higher energy excitations on the scale of a few tens to hundreds of eV, especially in the context of inner-shell spectroscopies\cite{Groot2008}, the core-electrons cannot be disregarded and the much denser real-space grids or higher kinetic energy of planewaves necessary to describe the highly-localized inner-shell orbitals in turn significantly increase the computational cost of such simulations. Adaptive non-uniform real-space grids\cite{Gygi1995} and FP-LAPW\cite{Dewhurst2004} methods afford possible means to circumvent this issue but an alternate approach is to employ a localized basis-set framework. A linear-combination of atomic-orbitals (LCAO)\cite{Martin2008} approach allows naturally for an efficient treatment of electronic states localized near the atomic nucleus but while potentially sacrificing some variational freedom in the  regions of low electron density away from nuclei especially when highly diffuse electronic states are involved. Nevertheless, such an approach might represent a worthwhile compromise in the context of theoretical simulations aimed at understanding inner-shell spectroscopies employing extreme ultraviolet (XUV)\cite{Schultz} or X-ray radiation\cite{Groot2008}. 

The advent of X-ray free electron lasers (FELs)\cite{Emma2010} and of high harmonic generation based attosecond XUV laser pulses\cite{Schultz} has led to the development of novel ultrafast spectroscopies that utilize inner-shell excitations to investigate electron dynamics on femtosecond and sub-femtosecond time scales\cite{Goulielmakis2010,Leone2014}. In particular, attosecond XUV spectroscopy has been extended to solid-state systems exploring for instance, the early-time dynamics of electrons excited across the band gap of semiconductor materials irradiated by intense few femtosecond near-infrared (NIR) pulses\cite{Schultze2014,Zurch2017}. More recently attosecond time-resolved core-exciton dynamics have also been investigated in a solid-state context\cite{Moulet2017}. As experimental capabilities utilizing short XUV or X-ray pulses advance further into areas such as non-linear X-ray spectroscopy, theoretical tools that can treat valence and core electron dynamics efficiently and on the same footing are necessary\cite{Zhang2015}. All-electron Gaussian-type orbital (GTO) basis-set implementations of RT-TDDFT based on the standard length-gauge have already been demonstrated\cite{Lopata2011} and have been utilized for simulating inner-shell spectroscopies in molecular systems\cite{Lopata2012}. However, to our knowledge, such GTO RT-TDDFT implementations have not yet been extended to include periodic boundary conditions and the velocity-gauge formalism for treating condensed-phase systems. It is in this scenario that we explore a velocity-gauge implementation of RT-TDDFT within an LCAO basis-set framework with the aim of assessing the level of agreement with real-space grid or FP-LAPW methods that such an approach can facilitate.  The VG-RT-TDDFT implementation described here is incorporated into a development version of the SIESTA\cite{Soler2002} code which provides a density functional theory\cite{Hohenberg1964,Kohn1965} (DFT) platform employing a numerical atomic orbital basis set in conjunction with norm-conserving pseudopotentials\cite{Martin2008}. Length gauge implementations of RT-TDDFT in unofficial versions of the SIESTA code have been previously described in the literature\cite{Tsolakidis2002,Takimoto2007}. The current velocity-gauge implementation builds upon a previous length-gauge implementation by Takimoto et al \cite{Takimoto2007} which was utilized for investigating the non-linear response properties of molecular systems. The remainder of this article is organized as follows. In section \ref{form}, the RT-TDDFT formalism underlying this work is outlined, followed in section~\ref{impl} by implementation details specific to SIESTA. Simulations on prototypical systems are presented in section \ref{rslt} while comparing the numerical results to real-space grid and FP-LAPW methods. Some considerations for future work and conclusions are outlined in section~\ref{conc}.

\section{Formalism}\label{form}
The velocity-gauge RT-TDDFT formalism implemented in this work is based on the one due to Bertsch et al\cite{Bertsch2000,Yabana2012}. In RT-TDDFT, the time-dependent Kohn-Sham (TDKS) equations 
\begin{equation}
\imath\hbar\frac{\partial}{\partial t}\psi_i(\overrightarrow{r},t)=\hat{H}_{KS}\psi_i(\overrightarrow{r},t)
\end{equation}
where $\hat{H}_{KS}$ and $\psi_i(\overrightarrow{r},t)$ are  the Kohn-Sham (KS) single-particle Hamiltonian and orbitals respectively, are integrated in the time-domain. Using the KS Hamiltonian in the length-gauge normally adopted with finite systems, the TDKS equations for electron dynamics take the form
\begin{align}
	&\imath\hbar\frac{\partial}{\partial t}\psi_i(\overrightarrow{r},t)=\nonumber\\
	&\left\lbrace \frac{\overrightarrow{p}^2}{2m} + \hat{V}_{ion}+\int d\overrightarrow{r}^\prime\frac{e^2}{|\overrightarrow{r}-\overrightarrow{r}^\prime|}n(\overrightarrow{r}^\prime,t) + V_{xc}[n(\overrightarrow{r},t)] + e \overrightarrow{E}.\overrightarrow{r}\right\rbrace \psi_i(\overrightarrow{r},t)
\end{align}     
where $\hat{V}_{ion}$ is the electron-ion interaction, the integral represents the Hartree potential and $V_{xc}[n(\overrightarrow{r},t)]$ is the exchange-correlation (XC) potential. The interaction of the electrons with an external electric field is given by the dipole coupling $e \overrightarrow{E}.\overrightarrow{r}$. $n(\overrightarrow{r},t)$ is the electron density obtained from the KS orbitals by $n(\overrightarrow{r},t)=\sum_{i}|\psi_i(\overrightarrow{r},t)|^2$. The velocity-gauge form of the equations suitable for infinite periodic systems is obtained through a gauge transformation involving the vector potential\cite{Bertsch2000,Yabana2012}
\begin{align}
	\overrightarrow{A}(t) &= -c \int^t \overrightarrow{E}(t^\prime) dt^\prime \\
	\psi_i(\overrightarrow{r},t) &= \mathrm{exp}{\left[ \dfrac{\imath e}{\hbar c} \overrightarrow{A}(t).\overrightarrow{r}\right]} \tilde{\psi}_i(\overrightarrow{r},t) 
\end{align}
yielding the velocity-gauge TDKS equations
\begin{align}\label{vgtdks}
	&\imath\hbar\frac{\partial}{\partial t}\tilde{\psi}_i(\overrightarrow{r},t)=\nonumber\\
	&\left\lbrace \frac{1}{2m}\left[  \overrightarrow{p} +\frac{e}{c}\overrightarrow{A}(t)\right]^2  + \hat{\tilde{V}}_{ion}+\int d\overrightarrow{r}^\prime \frac{e^2}{|\overrightarrow{r}-\overrightarrow{r}^\prime|}n(\overrightarrow{r}^\prime,t) + V_{xc}[n(\overrightarrow{r},t)] \right\rbrace \tilde{\psi}_i(\overrightarrow{r},t)
\end{align} 
wherein the vector potential $\overrightarrow{A}(t)$ appears in the kinetic term. In the case where non-local pseudopotentials are used, the gauge field also transforms the electron-ion interaction term to $\hat{\tilde{V}}_{ion}$ related to $\hat{V}_{ion}$ by
\begin{equation}\label{nlps}
\hat{\tilde{V}}_{ion} \tilde{\psi}_i(\overrightarrow{r},t)=\int d\overrightarrow{r}^\prime\mathrm{exp}{\left[-\dfrac{\imath e}{\hbar c} \overrightarrow{A}(t).\overrightarrow{r}\right]}~V_{ion}(\overrightarrow{r}^\prime,\overrightarrow{r})~\mathrm{exp}{\left[ \dfrac{\imath e}{\hbar c} \overrightarrow{A}(t).\overrightarrow{r}^\prime\right]}\tilde{\psi}_i(\overrightarrow{r}^\prime,t)
\end{equation} 
Although the length-gauge and velocity-gauge forms yield the same result for finite systems, the Hamiltonian in equation~\ref{vgtdks} is periodic for spatially uniform external electric fields, which allows a Bloch representation to be used in solving the velocity-gauge TDKS equations. Integrating equation~\ref{vgtdks} in time yields the time-dependent electron density $n(\overrightarrow{r},t)=\sum_{i}|\tilde{\psi}_i(\overrightarrow{r},t)|^2$ as well as the time-dependent current
\begin{equation}
\overrightarrow{I}(t)=-\frac{e}{\Omega}\int_{\Omega}d\overrightarrow{r}\overrightarrow{j}(\overrightarrow{r},t) 
\end{equation}
where the time-dependent current density
\begin{equation}
\overrightarrow{j}(\overrightarrow{r},t)=\sum_{i}\frac{e}{2m} \left\lbrace \tilde{\psi}^*_i(\overrightarrow{r},t) \overrightarrow{\pi}\tilde{\psi}_i(\overrightarrow{r},t) + c.c \right\rbrace
\end{equation}
features the generalized momentum
\begin{equation}
\overrightarrow{\pi} = \frac{m}{\imath\hbar} [\overrightarrow{r}, \hat{H}_{KS}] = -\imath\hbar\overrightarrow{\nabla} + \frac{e}{c}\overrightarrow{A}(t)+\frac{\imath m}{\hbar}\left[\tilde{V}_{ion},\overrightarrow{r} \right] 
\end{equation}
which accounts for the possible use of nonlocal pseudopotentials. Once the time-dependent density and current density are available frequency domain quantities can be accessed through Fourier transforms\cite{Marques2012a,Bertsch2000}. For instance, an impulsive field $\overrightarrow{A}(t)=\overrightarrow{A}_0\theta(t)$, where $\theta(t)$ is the Heaviside step function, can be applied and the  time-dependent current Fourier transformed to yield in the linear response regime the frequency dependent conductivity
\begin{equation}\label{fft}
\sigma_{ij}(\omega)=-\frac{c}{A_{0j}}\int^{T}~dt~ \mathrm{exp}(\imath\omega t)f(t)I_i(t)
\end{equation}
and the frequency dependent dielectric function
\begin{equation}\label{lreps}
\varepsilon(\omega)=1+\frac{4\pi\imath\sigma(\omega)}{\omega}
\end{equation}
In equation~\ref{fft}, $f(t)$ represents a filtering function inserted to avoid an abrupt cutoff of the integrand at end of the time-evolution period $T$.

\section{Implementation in SIESTA}\label{impl}
The KS Hamiltonian on the right hand side of equation~\ref{vgtdks} differs from its length-gauge counterpart in the structure of the kinetic and non-local pseudopotential terms. The Hartree and semi-local exchange-correlation (XC) terms are structurally unaffected by the velocity-gauge transformation and within the adiabatic approximation, can be calculated using standard procedures laid out for ground state simulations\cite{Soler2002}.  Expanding the kinetic term that now includes the time-dependent vector potential yields
\begin{equation}
\frac{1}{2m}\left[  \overrightarrow{p} +\frac{e}{c}\overrightarrow{A}(t)\right]^2 = \frac{1}{2m}\left[ -\hbar^2 \nabla^2  -2\frac{\imath \hbar e}{c} \overrightarrow{A}(t).\overrightarrow{\nabla} + \frac{e^2}{c^2} \overrightarrow{A}(t).\overrightarrow{A}(t)\right]
\end{equation}
In constructing the Hamiltonian matrix elements $H_{mn}=\langle m|H|n \rangle$ over SIESTA basis functions ($m,n$), we note that the first term on the right hand side involving $\nabla^2$ is simply the standard kinetic energy term. The second term in the expression requires evaluating matrix elements of the $\overrightarrow{\nabla}$ operator in the LCAO basis. These are also obtained using the standard reciprocal space scheme for two-center integrals in SIESTA\cite{Soler2002}. Matrix elements of the third term involving $\overrightarrow{A}(t).\overrightarrow{A}(t)$ reduce to expressions of the form $A^2(t)S$ where $S$ is the overlap matrix since $\overrightarrow{A}(t)$ is spatially homogenous within the present description. The vector potential entering the modified nonlocal pseudopotential operator in equation~\ref{nlps} gives rise to two-center integrals of the form
\begin{equation}
I_{mn}=\int d\overrightarrow{r} \phi_{m}^*(\overrightarrow{r}-\overrightarrow{R_{mn}})exp[\frac{\imath}{c}\overrightarrow{A}(t).\overrightarrow{r}]\phi_n(\overrightarrow{r})
\end{equation}
where $\overrightarrow{R}_{mn}$ is the vector connecting the nuclear coordinates of the two atom-centered basis functions $\phi_{m}$ and $\phi_n$. The reciprocal space two-center integral scheme within SIESTA is however not easily applicable in this instance because of the exponential function containing $\overrightarrow{A}(t).\overrightarrow{r}$ in the integrand. We therefore evaluate this expression in real space by using radial Gauss-Legendre and angular Lebedev-Laikov\cite{Lebedev1999} quadrature grids centered on one of the two atoms involved. We find in practice that a grid comprised of 140 radial and 110 angular points is sufficient to obtain converged energies on the order of $\sim$1 meV per atom.  In the velocity-gauge framework, we utilize the Bloch representation which allows for an efficient description of periodic systems in terms of unit cells due to the fact that the KS Hamiltonian remains diagonal in $\bm{k}$-space, and the TDKS equations at different $\bm{k}$-points are not coupled. Accordingly, complex Bloch wavefunctions are time evolved and a Brilloun zone integration is carried out to evaluate quantities such as the time dependent density and current. Nevertheless, the prescriptions laid out previously for numerical integration of the length gauge TDKS equations in SIESTA are portable to the present $\bm{k}$-space description once we replace the relevant quantities with their  Bloch counterparts. We expand the velocity-gauge Kohn-Sham wavefunctions $\tilde{\psi}_{i\bm{k}}$ over a static Bloch-LCAO basis set $\lbrace\phi_{m\bm{k}}\rbrace$
\begin{equation}
\tilde{\psi}_{i\bm{k}}(\overrightarrow{r},t)=\sum_{m} \phi_{m\bm{k}}(\overrightarrow{r}) c_{im}^{\bm{k}}(t) 
\end{equation}
so that the time dependence enters via the expansion coefficients  $c_{im}^{\bm{k}}$. Adopting atomic units, the TDKS equation~\ref{vgtdks}, is then recast in terms of the coefficient vector $c_i^{\bm{k}}$ as
\begin{equation}\label{ctdks}
\imath\frac{\partial c_i^{\bm{k}}}{\partial t} = S^{-1}_{\bm{k}}H_{\bm{k}}c_i^{\bm{k}}
\end{equation}
where $H_{\bm{k}}$  and $S_{\bm{k}}$ represent the Bloch-downfolded Hamiltonian and overlap matrices in the SIESTA basis.  Efficient unitary time evolution of the coefficient vectors $c_i^{\bm{k}}$ is carried out using the Crank-Nicholson\cite{Crank1947} scheme  whereby
\begin{equation}
c_i^{\bm{k}}(t + \Delta t) = \frac{1-\imath S^{-1}_{\bm{k}}H_{\bm{k}}(t)\Delta t/2}{1 + \imath S^{-1}_{\bm{k}}H_{\bm{k}}(t)\Delta t/2}c_i^{\bm{k}}(t) + O(\Delta t^2)
\end{equation}
Additionally, we utilize parallelism over $\bm{k}$-points through both distributed and shared memory parallelization in order to speed up the simulations on multiprocessor architectures.  

\section{Results}\label{rslt}
\subsection{Linear response and laser induced dynamics in bulk Silicon}\label{bSi}
With the aim of comparing the LCAO basis implementation of VG-RT-TDDFT within SIESTA with existing real-space implementations, we carry out a number of simulations on prototypical bulk Silicon investigating both linear response and strong laser field induced dynamics. As the real-space grid VG-RT-TDDFT implementation, we employ the well-established ARTED code\cite{A.Sato2014} which is a precursor to the SALMON project~\cite{salmon}, and has been used extensively for studying laser induced valence electron dynamics in solid state systems\cite{Yabana2012,Sato2015,Sommer2016}. For describing the valence electronic structure of bulk Si, we employ Neon-core norm-conserving pseudopotentials with a [Ne]$3s^2,3p^2$ configuration within both ARTED and SIESTA. The cubic 8-atom conventional unit cell of bulk Si with a lattice parameter of $a=10.26$ a.u is used in the simulations in conjunction with a $\Gamma$-centered 16x16x16 $\bm{k}$-point grid for Brillouin zone sampling. Exchange-correlation (XC) effects are treated at the level of the adiabatic local density approximation\cite{Marques2012a} (ALDA) employing the Perdew-Zunger~\cite{Perdew1981} form of the LDA~\cite{Kohn1965}. Within ARTED a 16x16x16 real-space mesh is used as the basis to represent the KS wavefunctions. In SIESTA on the other hand, we use a basis set of double-$\zeta$ quality featuring Si:$\{3s(2\zeta), 4s(2\zeta), 2p(2\zeta), 3p(2\zeta), 3d(2\zeta + polarization)\}$ functions in the basis, for a total of $27$ atomic orbital functions per Si atom. Basis function cut-off radii are consistent with a polarized atomic orbital energy shift~\cite{Soler2002} of $\sim$120 meV with the largest cutoff radii around $\sim7.1$ a.u. The real space mesh cutoff in SIESTA for calculating the Hartree and XC potentials is set to 200 Ry. A time-step of $0.08~\mathrm{a.u}$ ($1.93$ as) is employed in time-propagating the KS equations.

\begin{figure}[htbp]
	\centering
	\includegraphics[scale=0.5]{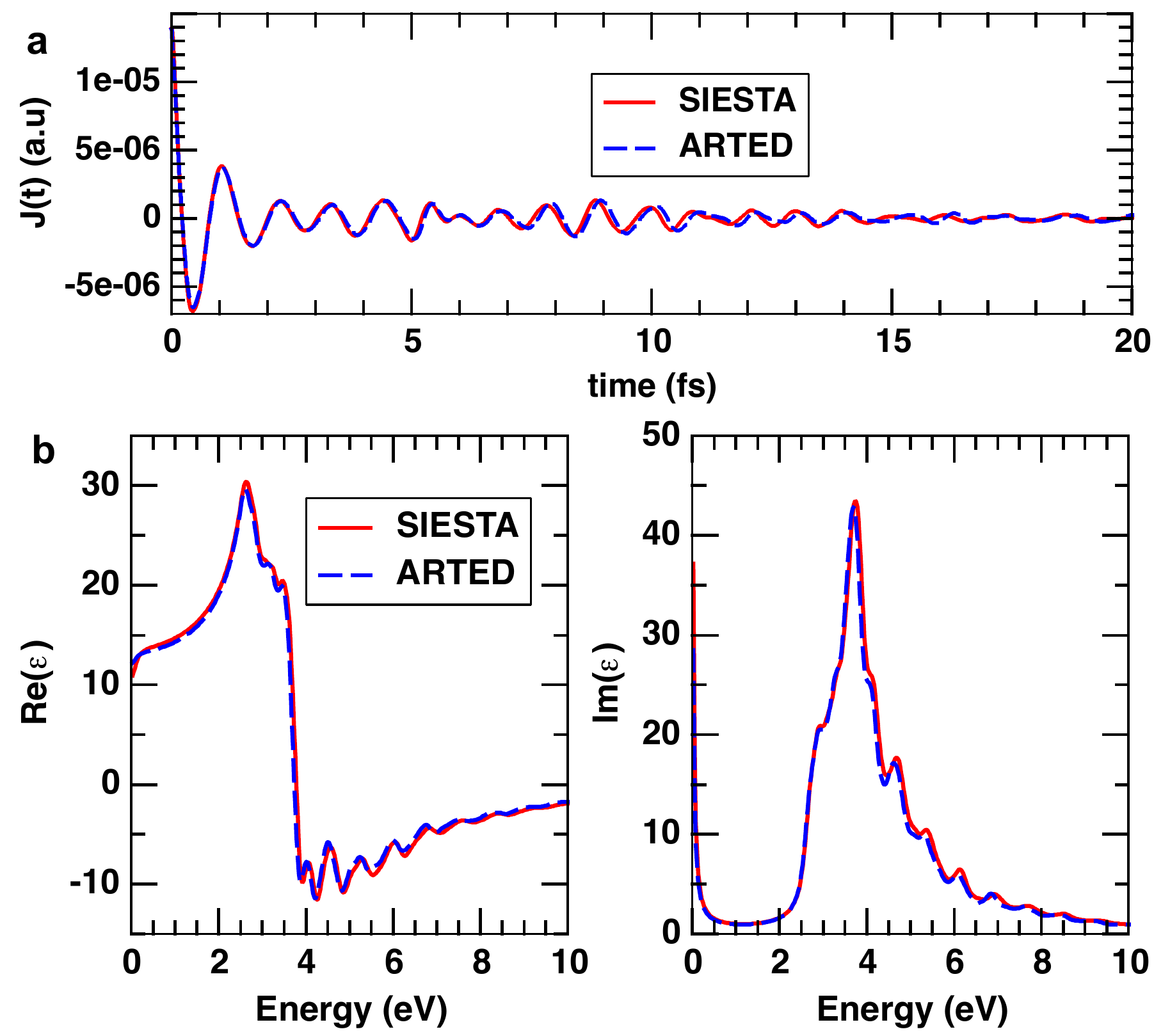}
	\caption{Linear dielectric response in bulk Si calculated from SIESTA and ARTED based VG-RT-TDDFT (a) Time-depended current induced by a weak impulsive electric field. (b) Real (left) and Imaginary (right) parts of the frequency dependent dielectric function $\epsilon(\omega)$}
	\label{lrsi}
\end{figure}

First we compare VG-RT-TDDFT results from SIESTA and ARTED for the linear dielectric response in bulk Si. Fig.~\ref{lrsi}(a) shows the time-dependent current $J(t)$ induced by a weak $0.001~\mathrm{a.u}$ electric field in the form of a $\delta$-function impulse applied along the $z$-axis at time zero. The currents from SIESTA and ARTED are closely matched in this instance. As outlined in equations~\ref{fft},\ref{lreps} the frequency dependent dielectric function $\epsilon(\omega)$ can be obtained after Fourier transforming $J(t)$. The real and imaginary parts of $\epsilon(\omega)$ from SIESTA and ARTED are plotted in Fig~\ref{lrsi}(b) and show good agreement. As discussed by Yabana et al~\cite{Yabana2012}, within the VG-RT-TDDFT approach a fictitious mode at zero frequency is observed which leads to a deviation from the correct analytical behaviour in $\epsilon(\omega)$ near $\omega=0$. This phenomenon is also observed in the SIESTA implementation and is consistent with ARTED. 

\begin{figure}[htbp]
	\centering
	\includegraphics[scale=0.39]{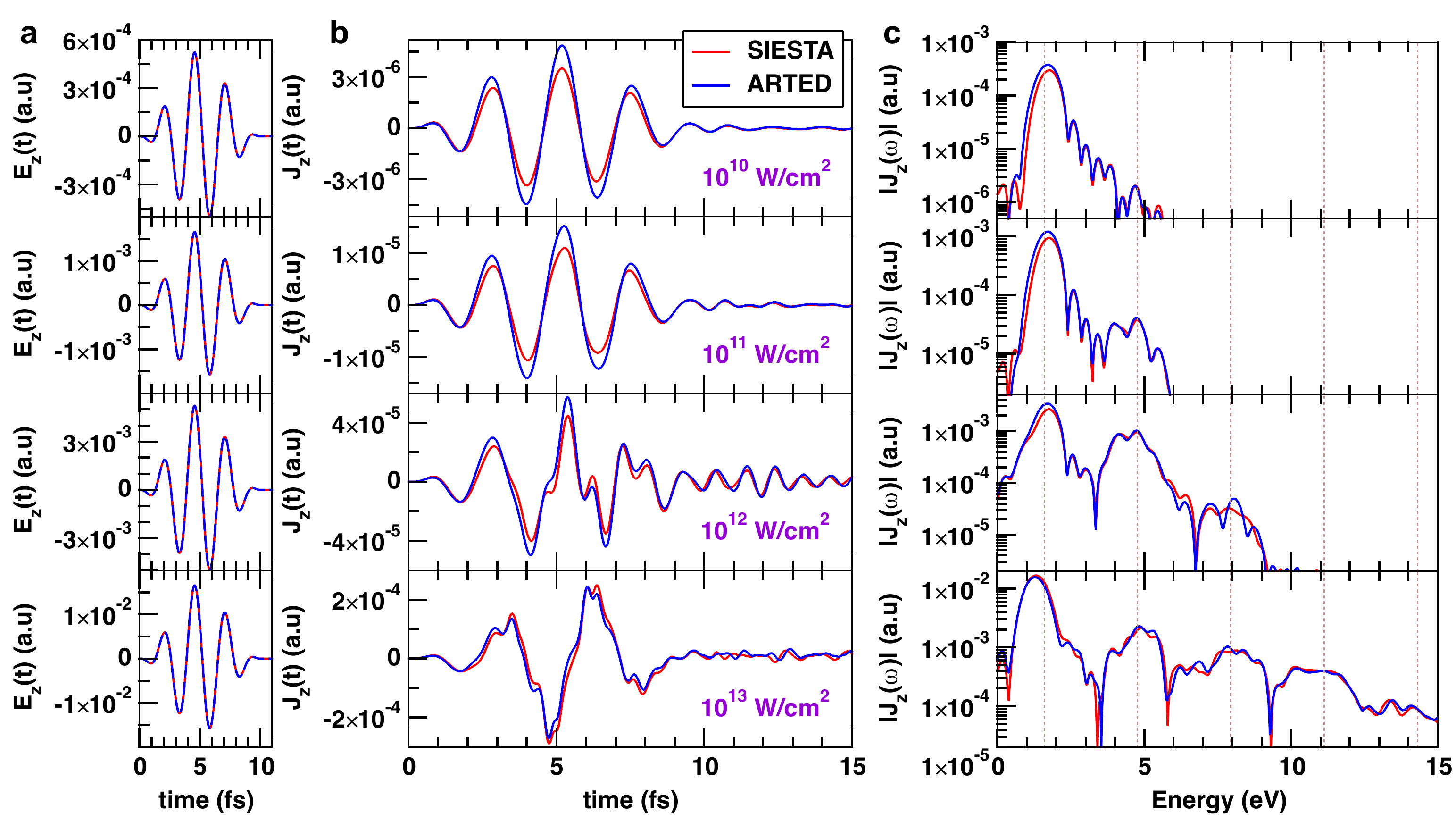}
	\caption{Interaction of a 10 fs infrared laser pulse of different intensities with bulk Si.  Red and blue curves correspond to SIESTA and ARTED respectively. (a) Magnitude and time-profile of the applied laser pulse for intensities ranging from $10^{10}$-$10^{13}$W/cm$^2$. (b) $z$-component of the time-dependent current induced by the applied laser field. (c) Frequency domain representation of the laser driven current indicating the production of higher harmonic oscillations with increasing laser intensity.}
	\label{lf-jvt}
\end{figure}

Next we investigate the interaction of intense  laser fields with bulk Si. For this purpose, we consider a 10 fs infrared (IR) laser pulse with a carrier frequency of 1.6 eV and a $\mathrm{sin}^2$ envelope centered at $t=5$ fs as shown in Fig.\ref{lf-jvt}(a). The electric field of the laser pulse is oriented along the $z$-axis of the conventional unitcell and we assume transverse boundary conditions so that surface polarization effects do not play a role~\cite{Yabana2012}. The action of this pulse for field intensities ranging from $10^{10}$ - $10^{13}$ $\mathrm{W/cm^2}$  (Fig.\ref{lf-jvt}(a)) is simulated within both SIESTA and ARTED. 

In Fig.~\ref{lf-jvt}(b) we plot the $z$-component of the laser induced current as a function of time. Overall, we find satisfactory agreement between SIESTA and ARTED for all of the considered field strengths. At early times, the calculated current $J_{z}(t)$ in SIESTA closely tracks its ARTED counterpart. Towards the middle of the pulse, the current oscillations in SIESTA near the field extrema are $\sim$10-20$\%$ weaker at low field intensities but the percentage magnitude of the deviation with respect to ARTED reduces at higher intensities. In the low intensity regime of $10^{10}-10^{11}~\mathrm{W/cm^2}$  the current $J_{z}(t)$ approximately mimics the time-profile of the applied field $E_{z}(t)$ albeit with a time-varying phase shift. At laser intensities over $10^{12} ~\mathrm{W/cm^2}$, additional oscillatory structures not present in $E_{z}(t)$ appear in $J_{z}(t)$. These correspond to nonlinear field-matter interactions, leading to the production of higher harmonics of the applied field frequency. In general, we find that the additional high-harmonic oscillations within $J_z(t)$ are well reproduced in SIESTA. A more quantitative comparison of the nonlinear effects is facilitated by Fourier transforming $J_z(t)$ and looking at the absolute magnitude of $J_z(\omega)$ in the frequency domain. This is shown in Fig.~\ref{lf-jvt}(c) for both SIESTA and ARTED.  Near the linear response regime at $10^{10}~\mathrm{W/cm^2}$, $|J_z(\omega)|$ exhibits only one prominent peak centered on the carrier frequency of the laser pulse at 1.6 eV. With increasing laser intensity, new peaks in $|J_z(\omega)|$ appear at odd harmonics of the carrier frequency and grow in strength. Even harmonics do not appear owing to the centrosymmetry of the Si lattice. At $10^{13}~\mathrm{W/cm^2}$  upto the $7^{th}$ harmonic can be readily identified. The agreement between SIESTA and ARTED in the frequency domain is also satisfactory over the range of intensities considered. We note in particular that  high-harmonic signals from the two codes compare well even though we do not employ very diffuse functions within the LCAO basis set. This is because in a solid state system like bulk Si, regions of extremely low electron density such as in the tails of molecular wavefunctions decaying into vacuum do not occur and the multiple-$\zeta$ basis is able to adequately describe density fluctuations.

\begin{figure}[htbp]
	\centering
	\includegraphics[scale=0.36]{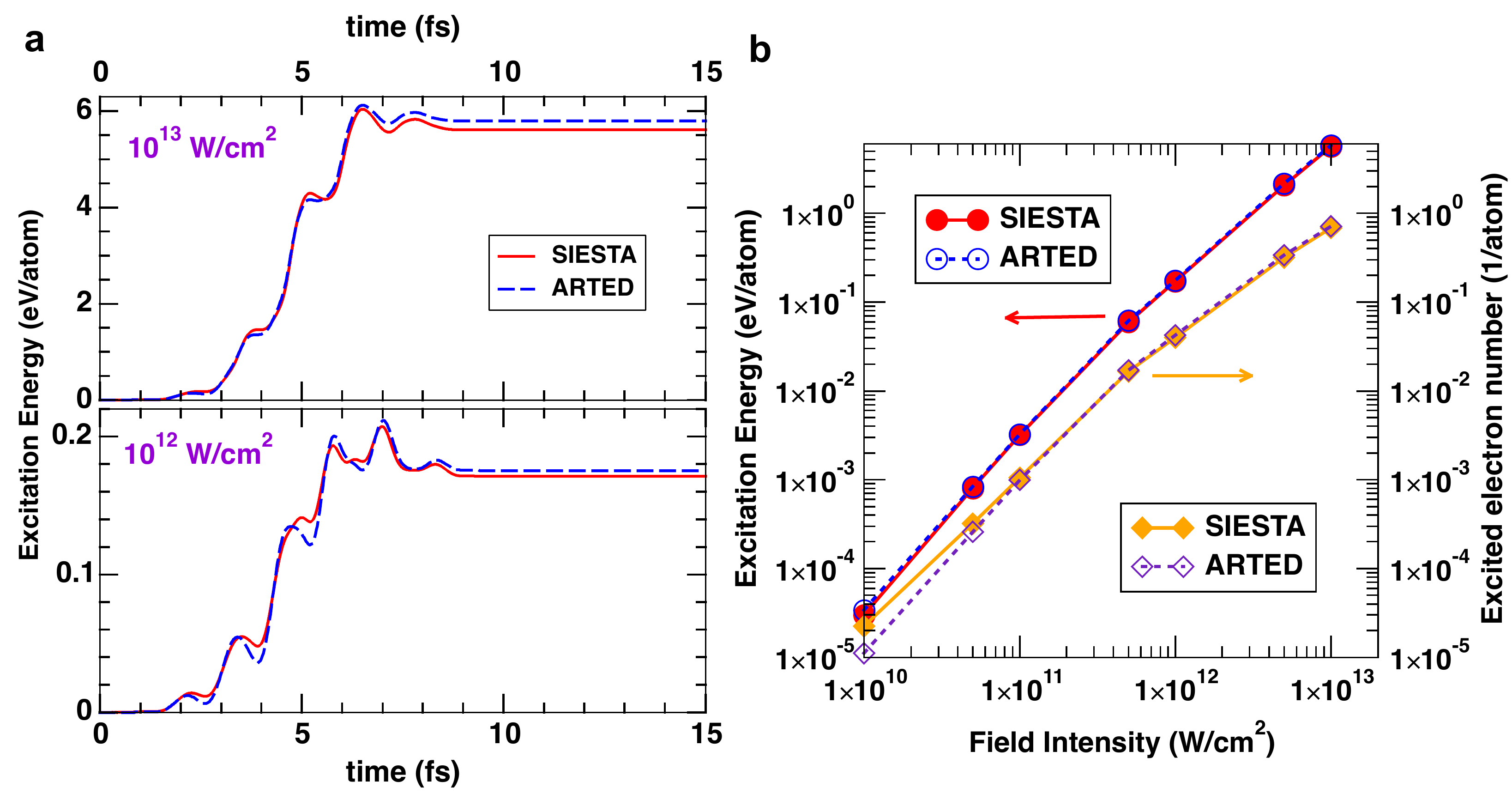}
	\caption{Energetics of laser-matter interaction in bulk Si for 10 fs infrared laser pulses of different intensities. (a) Excitation energy transferred to the material as function of interaction time with the laser pulse. Two different laser intensities $10^{13}$ W/cm$^2$ (top) and $10^{12}$ W/cm$^2$ (bottom) are considered. (b) Total excitation energy (left, circles) and total excited electron population (right, diamonds) in bulk Si at the end of interacting the laser pulse is plotted as a function of laser intensity.}
	\label{lf-evt}
\end{figure}
Having analyzed the laser driven current in both the time and frequency domains, we now consider the energetics of laser-matter interaction. As the laser pulse passes through bulk Si, electrons can be excited across the band gap leading to net energy absorption by the material. We note that, in the present instance, the central frequency of the exciting field at $\sim1.6$ eV, is well below the direct band gap ($\sim2.4$ eV in LDA) of Si. Therefore, excitation via single photon absorption is limited to the weak tail of the pulse's ($0.6$ eV FWHM) spectral distribution and multi-photon absorption or field induced tunneling processes are expected to contribute significantly to the excitation process\cite{Yabana2012}. Within VG-RT-TDDFT the total energy transferred to the material as a result of electronic excitation can be calculated as a function of time.  In Fig.~\ref{lf-evt}(a), we plot the time-dependent excitation energy from SIESTA and ARTED for two different laser intensities that are typical in laboratory pump-probe\cite{Schultze2014} ($10^{12}~\mathrm{W/cm^2}$ ) and high-harmonic generation\cite{Otobe2012,Floss2017} ($10^{13}~\mathrm{W/cm^2}$) experiments. The excitation energy increases with interaction time exhibiting oscillations that approximately coincide with half-cycles of the exciting field and remains constant after the pulse ends. As apparent from Fig.~\ref{lf-evt}(a) the same overall behavior is observed in SIESTA and ARTED. Near the center of the pulse, the amplitude of energy oscillations in SIESTA is slightly underestimated relative to ARTED especially at lower intensities. Nevertheless the total excitation energy per atom from SIESTA is within $5\%$ of the ARTED value regardless of the laser intensity. In Fig.~\ref{lf-evt}(b) we plot the excitation energy per atom (left) and the total number per atom of excited electrons in the conduction band (right) as a function of applied laser intensity. Very good agreement is observed between SIESTA and ARTED for both quantities across a range of experimentally relevant intensities.  
\begin{figure}[htbp]
	\centering
	\includegraphics[scale=0.5]{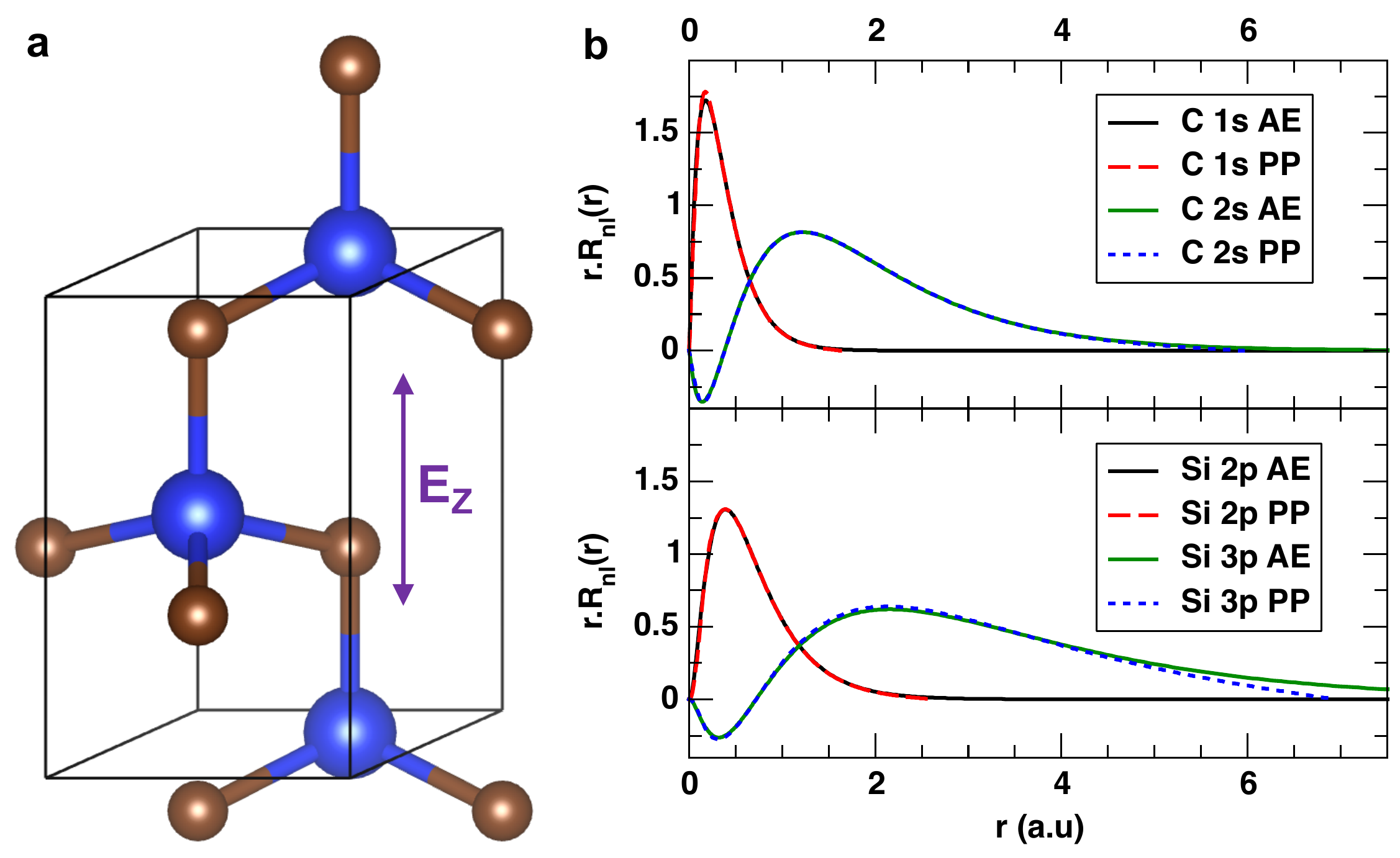}
	\caption{(a) Unitcell of the 2H polytype of SiC used for studying core-level linear response. The purple arrow indicates the direction of the applied electric field perturbation. (b) The radial function $r . R_{nl}(r)$ where $R_{nl}(r)$ is the solution of the radial Schrodinger equation is plotted as a function of radial distance ($r$) from the atomic nucleus. All-electron (AE) and pseudopotential (PP) quantities for the C $1s,2s$ (top) and Si $2p,3p$ (bottom) channels are compared. The PP functions correspond to the radial part of the first $\zeta$ basis function used in SIESTA. }
	\label{sic-ae}
\end{figure}
\subsection{Core-level spectroscopy in Silicon Carbide}\label{cls}
The primary convenience offered by an LCAO based RT-TDDFT approach over real-space grid basis methods is with respect to the treatment of localized core orbitals relevant to extreme ultraviolet (XUV) and X-ray spectroscopies. In this subsection we therefore investigate core-level response in $\mathrm{2H}$ Silicon Carbide (SiC) (Fig.~\ref{sic-ae}(a)) to demonstrate the utility of the present LCAO approach in this context. In particular we consider the L-edge of Si and the K-edge of C in the XUV and soft X-ray ranges respectively. Semi-core and core states can be treated within SIESTA by constructing appropriately modified pseudopotentials\cite{Martin2008}. For instance, in the case of the C K-edge (Si L-edge), we prepare a pseudopotential that explicitly pseudizes the C:$\{1s, 2p, 3d\}$ (Si:$\{2s,2p,3d\}$) states. The C $2s$ (Si $3p$) state is then automatically obtained as a higher energy solution of the atomic Schrodinger equation and its wavefunction has the correct nodal structure due to orthogonality with the C $1s$ (Si $2p$) wavefunction (See Fig.~\ref{sic-ae}(b)). The C $1s$ (Si $2p$) wavefunction is naturally consistent with the nodelessness criterion of standard pseudization techniques. Therefore, with an appropriate choice of the pseudopotential matching radius, both the C~$1s,2s$ (Si $2p,3p$) pseudo wavefunctions and eigenvalues closely resemble the corresponding all-electron counterparts as shown in Fig.~\ref{sic-ae}(b). Similar considerations apply for the $2s$ and $3s$ states of Si. Such a pseudopotential can be incorporated into the  SIESTA framework to model specific semi-core states in solid-state and molecular simulations. 
\begin{figure}[htbp]
	\centering
	\includegraphics[scale=0.56]{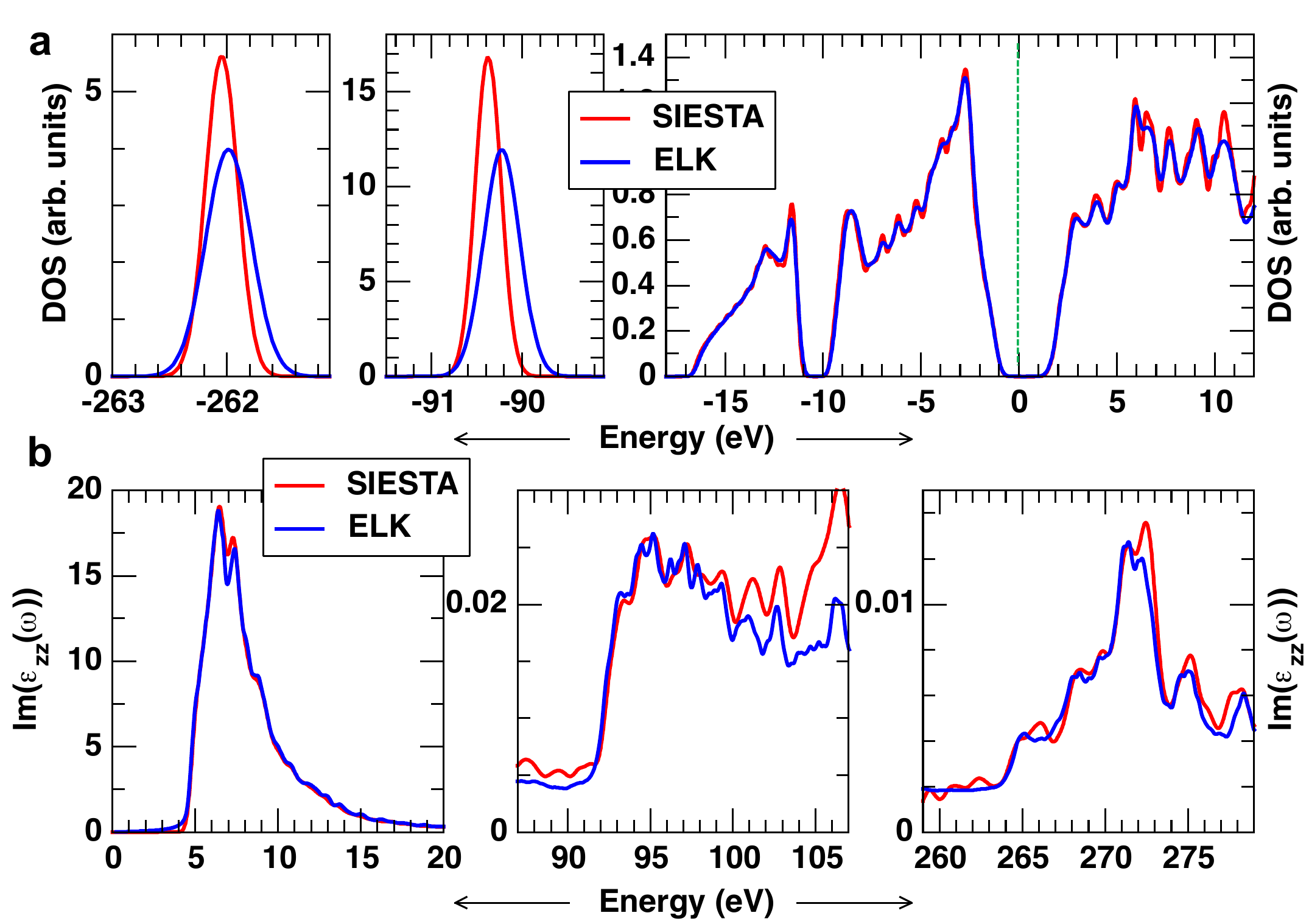}
	\caption{(a) Core and valence electronic density of states (DOS) in 2H SiC obtained from SIESTA (red) and Elk (blue). C $1s$ (left), Si $2p$ (middle) core-state and valence (right) DOS is shown. The dashed line (green) indicates the Fermi level (b) Imaginary part of the $zz$-component of the frequency dependent linear dielectric function in SiC calculated within the independent-particle approximation from SIESTA (red) and Elk (blue). Energy ranges corresponding to valence (left), Si L-edge (middle) and C K-edge (right) excitations are shown.}
	\label{sic-dos}
\end{figure}

We consider the 2H polytype of SiC which contains four atoms within the unitcell (Fig~\ref{sic-ae}(a)), in its experimentally determined geometry~\cite{wyckoff1963crystal} with lattice parameters of $\{\mathrm{a}=3.076\mathrm{\AA},\mathrm{c}=5.048\mathrm{\AA}\}$. We once again employ a double-$\zeta$ quality basis set featuring Si:$\{2s(2\zeta$), $3s(2\zeta)$, $4s(2\zeta)$, $2p(2\zeta)$, $3p(2\zeta)$, $3d(2\zeta)\}$,  C:$\{1s(2\zeta)$, $2s(2\zeta)$, $3s(2\zeta)$, $2p(2\zeta)$, $3d(2\zeta)\}$ functions in conjunction with a higher real space mesh cutoff of $400$ Ry for calculating the Hartree and XC potentials within SIESTA. Electronic structure results in this case are compared against those from the full-potential LAPW method implemented in the Elk~\cite{Dewhurst2004} code. Within Elk, a 24x24x40 planewave grid in conjunction with 28 local orbitals is utilized as the basis set. In both codes, a 20x20x12 $\bm{k}$-point grid is used for Brillouin zone sampling.  The electronic density of states (DOS) of SiC calculated from SIESTA and Elk are shown in Fig.~\ref{sic-dos}(a) over an extended energy range. As one would expect, very good agreement is obtained for the DOS in a range spanning several eV on either side of the Fermi level. For the Si $2p$ (C $1s$) core states near -91 eV (-262 eV), we find that KS energy band centers from the two codes agree to within 0.15 eV (0.06 eV) with the SIESTA eigenenergies slightly blue-shifted. We note that this difference on the order of 0.1\% is small compared to other sources of uncertainly within the theoretical approach such as due to the XC approximation. Furthermore, since spin-orbit coupling is neglected in the present simulations, no splitting of the Si $2p$ core-level DOS is seen. In Fig.~\ref{sic-dos}(b) the imaginary part of the frequency dependent dielectric function ($\epsilon_{zz}(\omega)$) as obtained from SIESTA and Elk, calculated using the independent-particle approximation (IPA) for light polarization along the $\mathrm{c}$-axis of the crystal (See Fig.~\ref{sic-ae}(a)) is compared over a 280 eV energy range. Spectral features originating from relevant low-energy valence and high-energy core excitations show satisfactory agreement both with respect to frequencies and relative oscillator strengths especially at energies near the absorption edges.  With regards to core-level spectra, IPA derived frequencies show small differences on the order of 0.1 eV consistent with those in the underlying KS eigenvalues.  Furthermore, while spectral features are  in good agreement overall, absolute oscillator strengths at higher energies are somewhat larger in the LCAO spectra at the Si L-edge.  We tentatively assign this to differences in the description of continuum wavefunctions in LCAO and FP-LAPW methods while noting that in the latter approach, oscillator strengths at high energies depend somewhat on the choice of muffin-tin radii.  These differences nevertheless should not affect spectral interpretation in the near-edge region.

\begin{figure}[htbp]
	\centering
	\includegraphics[scale=0.56]{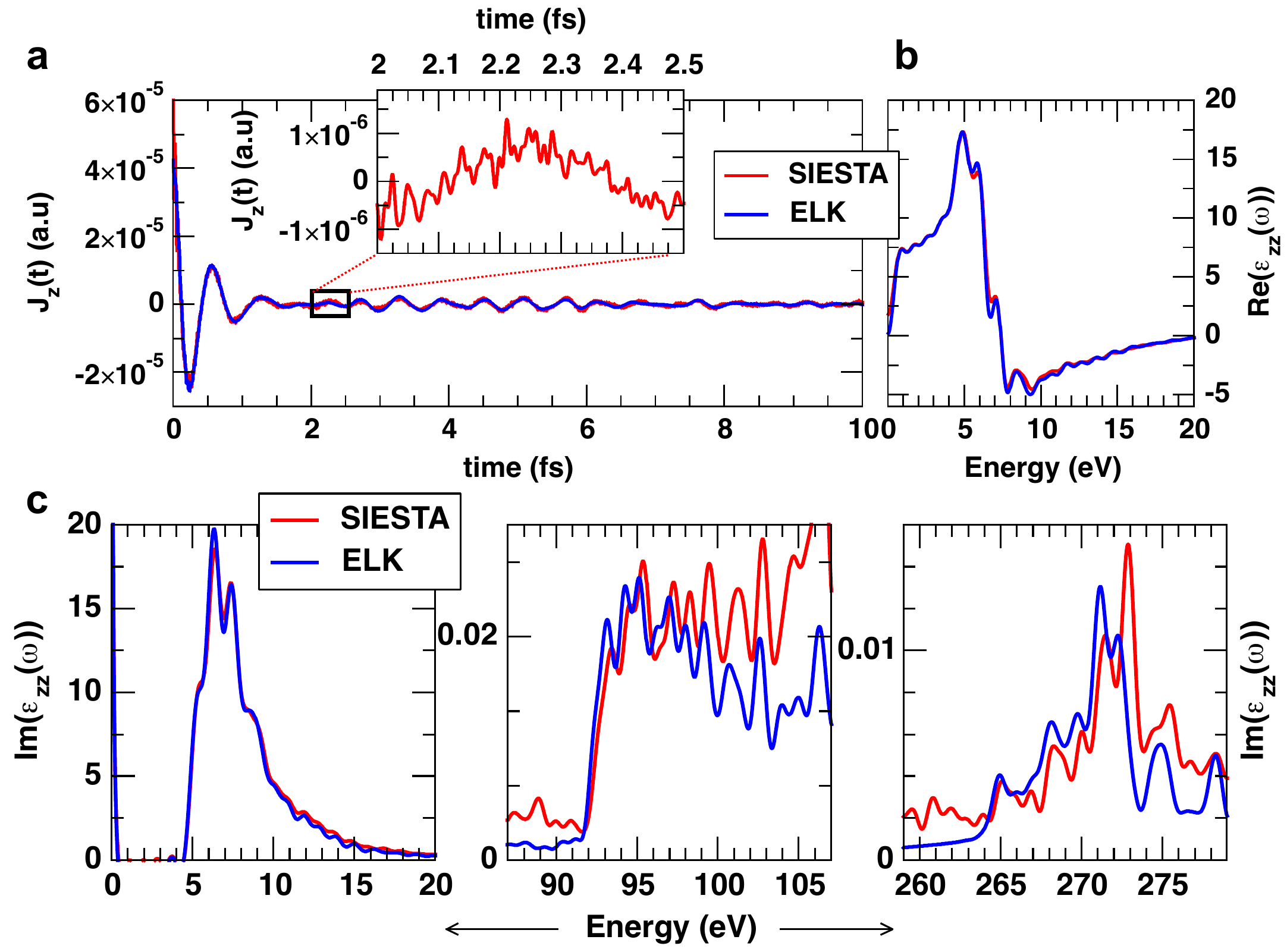}
	\caption{Linear dielectric response in 2H SiC calculated from SIESTA (red curves) and Elk (blue curves) based VG-RT-TDDFT (a) Time-depended current induced by a weak impulsive electric field. The inset shows high frequency core electron dynamics superposed on a slower valence oscillation. (b) Real part of the $zz$-component of the frequency dependent dielectric function ($\epsilon_{zz}(\omega)$) in the valence excitation region. (c) Imaginary part of $\epsilon_{zz}(\omega)$ plotted over energy ranges corresponding to valence (left), Si L-edge (middle) and C K-edge (right) excitations }
	\label{sic-td}
\end{figure}

\begin{figure}[htbp]
	\centering
	\includegraphics[scale=0.56]{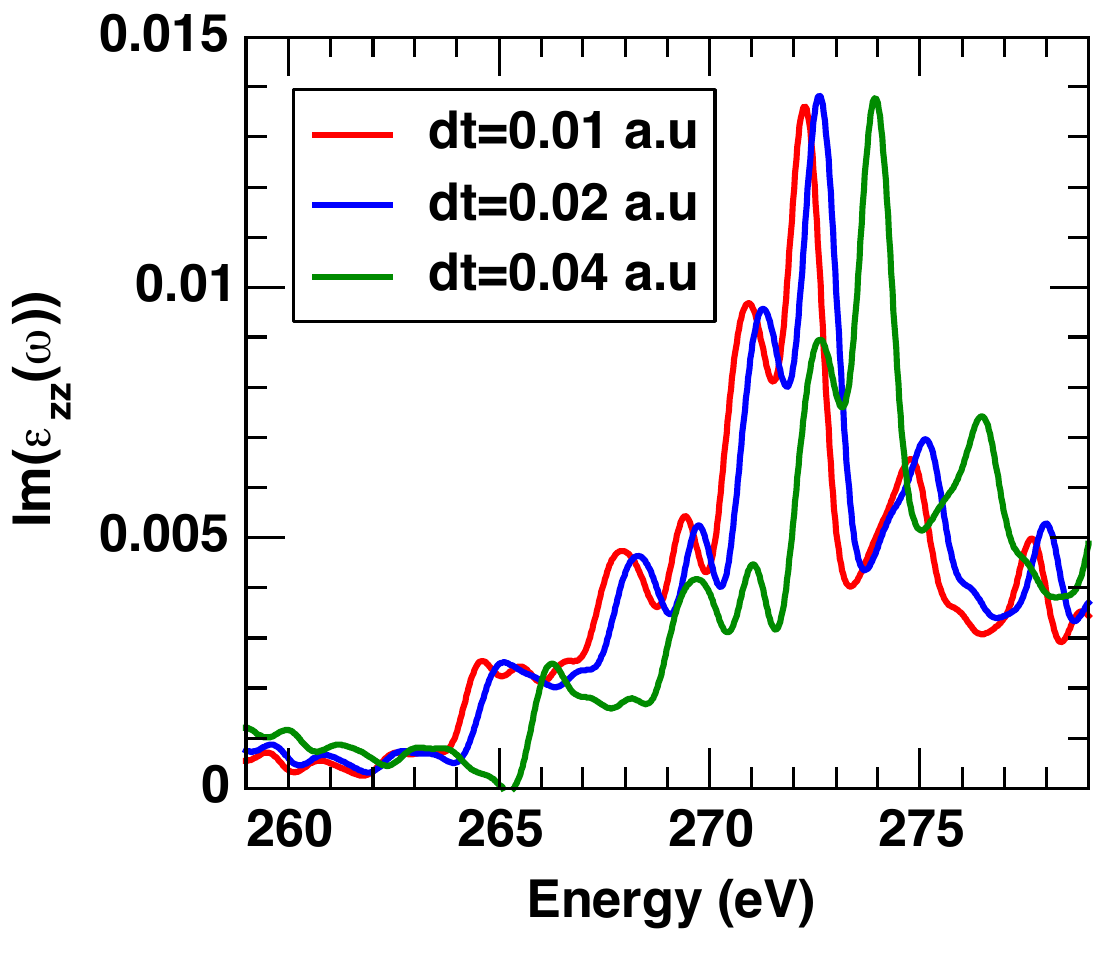}
	\caption{Convergence of the imaginary part of the $zz$-component of the linear dielectric function $\epsilon_{zz}(\omega)$ with integration time step ($dt$) used in the real-time simulation. Spectral features near the C K-edge are considered. }
	\label{sic-dt}
\end{figure}

Next we investigate real-time response in SiC using the VG-RT-TDDFT implementations in SIESTA and Elk with a particular emphasis on core-level excitations at the Si L- and C K-edges.  Relevant numerical parameters for real-time simulations are the same as the ones adopted to calculate the DOS and IPA response. Within Elk, a total of 115 KS states, with 18 occupied and 97 empty bands are included in the real-time description. Since core excitations at high energies are characterized by rapid oscillations of the electron density, a smaller time step is typically required to propagate the TDKS equations compared to valence-only simulations. Accordingly, we employ a time step of 0.01 a.u in both codes (See also discussion around Fig.~\ref{sic-dt}). By the same token however, time propagation does not need to be carried out for long periods to sample the fast core-electron oscillations, so the system is propagated in this instance for a total of 10 fs.  In Fig.~\ref{sic-td}, the time-dependent current density $J_z(t)$ induced in response to a weak 0.001 a.u impulsive electric field applied at time zero along the $\mathrm{c}$-axis of SiC is plotted. On the time scale relevant to valence oscillations, the SIESTA and Elk results for $J_z(t)$ are almost identical. The inset in Fig.~\ref{sic-td}(a) shows a magnified view of $J_z(t)$ from SIESTA during the interval between 2 - 2.5 fs from which it is apparent that the current density is characterized by high frequency core-electron oscillations of a smaller amplitude superimposed imposed upon slower valence oscillations. As before, $J_z(t)$ can be Fourier transformed to calculate the $zz$-component of the linear dielectric function $\epsilon_{zz}(\omega)$ in the frequency domain whose real and imaginary parts are plotted in Fig~\ref{sic-td}(b,c).  The real part (Fig~\ref{sic-td}) is plotted over a 20 eV range in the valence region and exhibits good agreement between the LCAO and FP-LAPW approaches. The imaginary part of $\epsilon_{zz}(\omega)$, which is relevant to absorption spectroscopies, is plotted in Fig~\ref{sic-td}(c) for both valence and core excitations spanning 20 eV energy ranges in each case. Once again good agreement is observed in the valence region between LCAO and FP-LAPW methods. Near the core-excitation edges, the energy positions of spectral peaks in the X-ray absorption fine structure exhibit satisfactory agreement over the considered energy range.  As in the case of the IPA spectra in Fig~\ref{sic-dos}, absolute  intensities in the core-level spectra are seen to be larger in the LCAO calculation at higher energies above the edge. Furthermore, we note that especially at the C K-edge, the contribution to the overall spectrum from a decaying background of high-energy valence excitations is different in the two codes. In SIESTA the high-energy background contribution leads to some additional modulations which are apparent below the absorption edge. These features that in reality would be washed out by lifetime broadening effects, show up as oscillatory features in the real-time spectra obtained via Fourier transforms. The Elk spectrum is free of such features since only empty states spanning approximately $100$ eV above the Fermi level are effectively included in the description.  These differences are not crucial however, since the valence-excitation background at X-ray energies is not very sensitive to chemical differences and in the context of time-resolved spectroscopies, does not contribute significantly to the differential pump-probe signal. 

Finally, we discuss briefly the issue of convergence of the core-level spectra with integration time step. In Fig.~\ref{sic-dt} we show the C K-edge spectra obtained from SIESTA based VG-RT-TDDFT for three different integration time steps between 0.01 - 0.04 a.u. At all three time steps, the Crank-Nicholson integration scheme is stable over the 10 fs propagation time. At the smallest time step of 0.01 a.u the absolute energy positions of the spectral features are converged to within 0.1 eV suggesting that the same is adequate to sample rapid oscillations due to C K-edge derived excitations. In contrast the spectra for time steps of 0.02 a.u and 0.04 a.u are shifted to higher energies. Nevertheless, it is apparent that while the lower sampling rate affects absolute energy positions, the relative energy differences between different spectral features are practically identical for all three sampling rates. This provides a numerical illustration of the approximate separability of the electron dynamics into fast core and slow valence oscillations with the former primarily determining the absolute energy positions of the core-excitation edges and the latter being more relevant for the detailed structure of the near-edge features. Therefore, in some applications where reproducing the absolute core-excitation edge energy is not crucial, larger time steps can in principle be used to simulate near-edge spectral features.  A similar separation of time scales argument has also been utilized by Lee et al for the simulation of core-level spectra within the local time-correlation approach~\cite{Lee2012}.  

\section{Conclusions and Outlook}\label{conc}
In summary, we have implemented the velocity-gauge formalism of real-time TDDFT within a numerical atomic orbital based first-principles framework and carried out a comparative assessment of time-varying electric field response properties against results from real-space grid (RSG) and FP-LAPW codes. The overall agreement between the current LCAO basis approach and RSG or FP-LAPW methods is satisfactory in that the basis set or pseudopotential driven differences in the observable quantities simulated are small compared to errors intrinsic to the underlying first-principles density functional approximations. LCAO and RSG/FP-LAPW predicted excitation frequencies differ negligibly in the valence region and by $\sim$0.1 eV in the core-region. The energetics of laser-matter interaction are also well described by the LCAO framework and are within $5\%$ of RSG results. The fact that this level of agreement is achieved without employing very large basis sets or targeted optimization of the basis functions suggests that there is potentially scope for obtaining even closer agreement where necessary, with some additional effort. The primary advantages afforded by the LCAO approach are the relatively small size and short-ranged character of the basis set which leads to efficient representations for the density and Hamiltonian matrices and favorbale scaling with increasing system size~\cite{Soler2002}.  The present LCAO implementation of VG-RT-TDDFT in-principle retains these advanages while enabling the study of time-resolved and nonlinear XUV/X-ray spectroscopies at similar computational cost as simulating valence-only dynamics. However, in the context of modeling laser-matter interaction in small unitcells with dense reciprocal space-sampling while employing adiabatic semi-local XC functionals, it does not afford significant advantages over more mature existing FP-LAPW based VG-RT-TDDFT\cite{Krieger2015}. While some efficiencies could potentially be harnessed in supercell simulations of low dimensional systems that include large vacuum regions, in order to fully take advantage of the small basis set sizes inherent to the LCAO approach, the implementation must be extended to make use of non-local XC functionals~\cite{Sato2015a} to describe excitonic effects and density-matrix evolution to incorporate coupling to external baths within a Liouville picture~\cite{Marques2012a,Burke2005}. Efforts along these lines are currently underway.

\section*{Acknowledgements}
The work of CDP, JJR, FDV and JJK  is carried out within the Theory Institute for Materials and Energy Spectroscopies (TIMES) at SLAC, supported by the U.S. DOE, Office of Basic Energy Sciences, Division of Materials Sciences and Engineering, under contract no. DE-AC02-76SF00515. Early development work was performed by C.D.P and D.P as part of a User Project at The Molecular Foundry (TMF), LBNL, supported by the Office of Science, Office of Basic Energy Sciences, of the U.S. DOE, under contract no. DE-AC02-05CH11231, and by JJR, FDV and JJK at U. Washington under DOE BES Grant DE-FG02-ER45623.  Numerical simulations were executed on the Etna, Vulcan, Mako, and Lawrencium compute clusters, administered by the High-Performance Computing Services Group at LBNL. K.Y. is supported by JSPS KAKENHI 15H03674 and by JST CREST Grant Number JPMJCR16N5, Japan.

\section*{References}

\bibliography{VG-RT-TDDFT.bib}

\begin{thebibliography}{10}
\expandafter\ifx\csname url\endcsname\relax
  \def\url#1{\texttt{#1}}\fi
\expandafter\ifx\csname urlprefix\endcsname\relax\def\urlprefix{URL }\fi
\expandafter\ifx\csname href\endcsname\relax
  \def\href#1#2{#2} \def\path#1{#1}\fi

\bibitem{Runge1984}
E.~Runge, E.~K.~U. Gross,
  \href{http://link.aps.org/doi/10.1103/PhysRevLett.52.997}{{Density-Functional
  Theory for Time-Dependent Systems}}, Physical Review Letters 52~(12) (1984)
  997--1000.
\newblock \href {http://dx.doi.org/10.1103/PhysRevLett.52.997}
  {\path{doi:10.1103/PhysRevLett.52.997}}.
\newline\urlprefix\url{http://link.aps.org/doi/10.1103/PhysRevLett.52.997}

\bibitem{Marques2012a}
M.~A. Marques, N.~T. Maitra, F.~M. Nogueira, E.~Gross, A.~Rubio (Eds.),
  \href{http://link.springer.com/10.1007/978-3-642-23518-4}{{Fundamentals of
  Time-Dependent Density Functional Theory}}, Vol. 837 of Lecture Notes in
  Physics, Springer Berlin Heidelberg, Berlin, Heidelberg, 2012.
\newblock \href {http://dx.doi.org/10.1007/978-3-642-23518-4}
  {\path{doi:10.1007/978-3-642-23518-4}}.
\newline\urlprefix\url{http://link.springer.com/10.1007/978-3-642-23518-4}

\bibitem{Ullrich2012}
C.~Ullrich, {Time-dependent density-functional theory : concepts and
  applications}, Oxford University Press, 2012.

\bibitem{Bertsch2000}
G.~F. Bertsch, J.~I. Iwata, A.~Rubio, K.~Yabana,
  \href{https://link.aps.org/doi/10.1103/PhysRevB.62.7998}{{Real-space,
  real-time method for the dielectric function}}, Physical Review B - Condensed
  Matter and Materials Physics 62~(12) (2000) 7998--8002.
\newblock \href {http://arxiv.org/abs/0005512v1} {\path{arXiv:0005512v1}},
  \href {http://dx.doi.org/10.1103/PhysRevB.62.7998}
  {\path{doi:10.1103/PhysRevB.62.7998}}.
\newline\urlprefix\url{https://link.aps.org/doi/10.1103/PhysRevB.62.7998}

\bibitem{Tsolakidis2002}
A.~Tsolakidis, D.~S{\'{a}}nchez-Portal, R.~M. Martin,
  \href{https://link.aps.org/doi/10.1103/PhysRevB.66.235416}{{Calculation of
  the optical response of atomic clusters using time-dependent density
  functional theory and local orbitals}}, Physical Review B 66~(23) (2002)
  235416.
\newblock \href {http://arxiv.org/abs/0109488} {\path{arXiv:0109488}}, \href
  {http://dx.doi.org/10.1103/PhysRevB.66.235416}
  {\path{doi:10.1103/PhysRevB.66.235416}}.
\newline\urlprefix\url{https://link.aps.org/doi/10.1103/PhysRevB.66.235416}

\bibitem{Takimoto2007}
Y.~Takimoto, F.~D. Vila, J.~J. Rehr,
  \href{http://aip.scitation.org/doi/10.1063/1.2790014}{{Real-time
  time-dependent density functional theory approach for frequency-dependent
  nonlinear optical response in photonic molecules}}, Journal of Chemical
  Physics 127~(15) (2007) 154114.
\newblock \href {http://dx.doi.org/10.1063/1.2790014}
  {\path{doi:10.1063/1.2790014}}.
\newline\urlprefix\url{http://aip.scitation.org/doi/10.1063/1.2790014}

\bibitem{Meng2008}
S.~Meng, E.~Kaxiras,
  \href{http://aip.scitation.org/doi/10.1063/1.2960628}{{Real-time, local
  basis-set implementation of time-dependent density functional theory for
  excited state dynamics simulations}}, Journal of Chemical Physics 129~(5)
  (2008) 054110.
\newblock \href {http://dx.doi.org/10.1063/1.2960628}
  {\path{doi:10.1063/1.2960628}}.
\newline\urlprefix\url{http://aip.scitation.org/doi/10.1063/1.2960628}

\bibitem{Lopata2011}
K.~Lopata, N.~Govind,
  \href{http://pubs.acs.org/doi/abs/10.1021/ct200137z}{{Modeling Fast Electron
  Dynamics with Real-Time Time-Dependent Density Functional Theory: Application
  to Small Molecules and Chromophores}}, Journal of Chemical Theory and
  Computation 7~(5) (2011) 1344--1355.
\newblock \href {http://dx.doi.org/10.1021/ct200137z}
  {\path{doi:10.1021/ct200137z}}.
\newline\urlprefix\url{http://pubs.acs.org/doi/abs/10.1021/ct200137z}

\bibitem{Castro2012}
A.~Castro, J.~Werschnik, E.~K.~U. Gross,
  \href{https://link.aps.org/doi/10.1103/PhysRevLett.109.153603}{{Controlling
  the dynamics of many-electron systems from first principles: A combination of
  optimal control and time-dependent density-functional theory}}, Physical
  Review Letters 109~(15) (2012) 153603.
\newblock \href {http://dx.doi.org/10.1103/PhysRevLett.109.153603}
  {\path{doi:10.1103/PhysRevLett.109.153603}}.
\newline\urlprefix\url{https://link.aps.org/doi/10.1103/PhysRevLett.109.153603}

\bibitem{Andrade2012}
X.~Andrade, J.~Alberdi-Rodriguez, D.~A. Strubbe, M.~J.~T. Oliveira,
  F.~Nogueira, A.~Castro, J.~Muguerza, A.~Arruabarrena, S.~G. Louie,
  A.~Aspuru-Guzik, A.~Rubio, M.~A.~L. Marques,
  \href{http://iopscience.iop.org/0953-8984/24/23/233202/article/}{{Time-dependent
  density-functional theory in massively parallel computer architectures: the
  OCTOPUS project.}}, Journal of physics. Condensed matter : an Institute of
  Physics journal 24~(23) (2012) 233202.
\newblock \href {http://dx.doi.org/10.1088/0953-8984/24/23/233202}
  {\path{doi:10.1088/0953-8984/24/23/233202}}.
\newline\urlprefix\url{http://iopscience.iop.org/0953-8984/24/23/233202/article/}

\bibitem{Yabana2012}
K.~Yabana, T.~Sugiyama, Y.~Shinohara, T.~Otobe, G.~F. Bertsch,
  \href{http://link.aps.org/doi/10.1103/PhysRevB.85.045134}{{Time-dependent
  density functional theory for strong electromagnetic fields in crystalline
  solids}}, Physical Review B 85~(4) (2012) 045134.
\newblock \href {http://dx.doi.org/10.1103/PhysRevB.85.045134}
  {\path{doi:10.1103/PhysRevB.85.045134}}.
\newline\urlprefix\url{http://link.aps.org/doi/10.1103/PhysRevB.85.045134}

\bibitem{Wang2013}
R.~Wang, D.~Hou, X.~Zheng,
  \href{https://link.aps.org/doi/10.1103/PhysRevB.88.205126}{{Time-dependent
  density-functional theory for real-time electronic dynamics on material
  surfaces}}, Physical Review B - Condensed Matter and Materials Physics
  88~(20) (2013) 205126.
\newblock \href {http://arxiv.org/abs/1307.5762} {\path{arXiv:1307.5762}},
  \href {http://dx.doi.org/10.1103/PhysRevB.88.205126}
  {\path{doi:10.1103/PhysRevB.88.205126}}.
\newline\urlprefix\url{https://link.aps.org/doi/10.1103/PhysRevB.88.205126}

\bibitem{Krieger2015}
K.~Krieger, J.~K. Dewhurst, P.~Elliott, S.~Sharma, E.~K. Gross,
  \href{http://pubs.acs.org/doi/10.1021/acs.jctc.5b00621}{{Laser-Induced
  Demagnetization at Ultrashort Time Scales: Predictions of TDDFT}}, Journal of
  Chemical Theory and Computation 11~(10) (2015) 4870--4874.
\newblock \href {http://dx.doi.org/10.1021/acs.jctc.5b00621}
  {\path{doi:10.1021/acs.jctc.5b00621}}.
\newline\urlprefix\url{http://pubs.acs.org/doi/10.1021/acs.jctc.5b00621}

\bibitem{Goings2016}
J.~J. Goings, X.~Li, \href{http://aip.scitation.org/doi/10.1063/1.4953668}{{An
  atomic orbital based real-time time-dependent density functional theory for
  computing electronic circular dichroism band spectra}}, The Journal of
  Chemical Physics 144~(23) (2016) 234102.
\newblock \href {http://dx.doi.org/10.1063/1.4953668}
  {\path{doi:10.1063/1.4953668}}.
\newline\urlprefix\url{http://aip.scitation.org/doi/10.1063/1.4953668}

\bibitem{Nguyen2016}
T.~S. Nguyen, J.~H. Koh, S.~Lefelhocz, J.~Parkhill,
  \href{http://pubs.acs.org/doi/abs/10.1021/acs.jpclett.6b00421}{{Black-Box,
  Real-Time Simulations of Transient Absorption Spectroscopy}}, Journal of
  Physical Chemistry Letters 7~(8) (2016) 1590--1595.
\newblock \href {http://dx.doi.org/10.1021/acs.jpclett.6b00421}
  {\path{doi:10.1021/acs.jpclett.6b00421}}.
\newline\urlprefix\url{http://pubs.acs.org/doi/abs/10.1021/acs.jpclett.6b00421}

\bibitem{Provorse2016}
M.~R. Provorse, C.~M. Isborn,
  \href{http://doi.wiley.com/10.1002/qua.25096}{{Electron dynamics with
  real-time time-dependent density functional theory}} (may 2016).
\newblock \href {http://dx.doi.org/10.1002/qua.25096}
  {\path{doi:10.1002/qua.25096}}.
\newline\urlprefix\url{http://doi.wiley.com/10.1002/qua.25096}

\bibitem{Yost2017}
D.~C. Yost, Y.~Yao, Y.~Kanai,
  \href{https://link.aps.org/doi/10.1103/PhysRevB.96.115134}{{Examining
  real-time time-dependent density functional theory nonequilibrium simulations
  for the calculation of electronic stopping power}}, Physical Review B 96~(11)
  (2017) 115134.
\newblock \href {http://dx.doi.org/10.1103/PhysRevB.96.115134}
  {\path{doi:10.1103/PhysRevB.96.115134}}.
\newline\urlprefix\url{https://link.aps.org/doi/10.1103/PhysRevB.96.115134}

\bibitem{Sato2015}
S.~A. Sato, K.~Yabana, Y.~Shinohara, T.~Otobe, K.~M. Lee, G.~F. Bertsch,
  \href{https://link.aps.org/doi/10.1103/PhysRevB.92.205413}{{Time-dependent
  density functional theory of high-intensity short-pulse laser irradiation on
  insulators}}, Physical Review B - Condensed Matter and Materials Physics
  92~(20) (2015) 205413.
\newblock \href {http://arxiv.org/abs/1412.1445} {\path{arXiv:1412.1445}},
  \href {http://dx.doi.org/10.1103/PhysRevB.92.205413}
  {\path{doi:10.1103/PhysRevB.92.205413}}.
\newline\urlprefix\url{https://link.aps.org/doi/10.1103/PhysRevB.92.205413}

\bibitem{Floss2017}
I.~Floss, G.~Wachter, C.~Lemell, S.~Sato, X.-M. Tong, K.~Yabana,
  J.~Burgd{\"{o}}rfer,
  \href{http://stacks.iop.org/1742-6596/875/i=5/a=042007?key=crossref.b12752297160a9c9fe5bad8ffccb004d}{{Simulation
  of High Harmonic Generation in Solids}}, Journal of Physics: Conference
  Series 875~(5) (2017) 042007.
\newblock \href {http://dx.doi.org/10.1088/1742-6596/875/5/042007}
  {\path{doi:10.1088/1742-6596/875/5/042007}}.
\newline\urlprefix\url{http://stacks.iop.org/1742-6596/875/i=5/a=042007?key=crossref.b12752297160a9c9fe5bad8ffccb004d}

\bibitem{Tancogne-Dejean2017}
N.~Tancogne-Dejean, O.~D. M{\"{u}}cke, F.~X. K{\"{a}}rtner, A.~Rubio,
  \href{https://link.aps.org/doi/10.1103/PhysRevLett.118.087403}{{Impact of the
  Electronic Band Structure in High-Harmonic Generation Spectra of Solids}},
  Physical Review Letters 118~(8) (2017) 087403.
\newblock \href {http://arxiv.org/abs/1609.09298} {\path{arXiv:1609.09298}},
  \href {http://dx.doi.org/10.1103/PhysRevLett.118.087403}
  {\path{doi:10.1103/PhysRevLett.118.087403}}.
\newline\urlprefix\url{https://link.aps.org/doi/10.1103/PhysRevLett.118.087403}

\bibitem{DeGiovannini2013}
U.~{De Giovannini}, G.~Brunetto, A.~Castro, J.~Walkenhorst, A.~Rubio,
  \href{http://doi.wiley.com/10.1002/cphc.201201007
  http://www.ncbi.nlm.nih.gov/pubmed/23520148}{{Simulating pump-probe
  photoelectron and absorption spectroscopy on the attosecond timescale with
  time-dependent density functional theory.}}, Chemphyschem : a European
  journal of chemical physics and physical chemistry 14~(7) (2013) 1363--76.
\newblock \href {http://dx.doi.org/10.1002/cphc.201201007}
  {\path{doi:10.1002/cphc.201201007}}.
\newline\urlprefix\url{http://doi.wiley.com/10.1002/cphc.201201007
  http://www.ncbi.nlm.nih.gov/pubmed/23520148}

\bibitem{Sommer2016}
A.~Sommer, E.~M. Bothschafter, S.~A. Sato, C.~Jakubeit, T.~Latka,
  O.~Razskazovskaya, H.~Fattahi, M.~Jobst, W.~Schweinberger, V.~Shirvanyan,
  V.~S. Yakovlev, R.~Kienberger, K.~Yabana, N.~Karpowicz, M.~Schultze,
  F.~Krausz,
  \href{http://www.nature.com/doifinder/10.1038/nature17650}{{Attosecond
  nonlinear polarization and light–matter energy transfer in solids}}, Nature
  534~(7605) (2016) 86--90.
\newblock \href {http://dx.doi.org/10.1038/nature17650}
  {\path{doi:10.1038/nature17650}}.
\newline\urlprefix\url{http://www.nature.com/doifinder/10.1038/nature17650}

\bibitem{Lucchini2016}
M.~Lucchini, S.~A. Sato, A.~Ludwig, J.~Herrmann, M.~Volkov, L.~Kasmi,
  Y.~Shinohara, K.~Yabana, L.~Gallmann, U.~Keller,
  \href{http://science.sciencemag.org/content/353/6302/916
  http://www.ncbi.nlm.nih.gov/pubmed/27563093}{{No Title}} 353~(6302) (2016)
  916--9.
\newblock \href {http://dx.doi.org/10.1126/science.aag1268}
  {\path{doi:10.1126/science.aag1268}}.
\newline\urlprefix\url{http://science.sciencemag.org/content/353/6302/916
  http://www.ncbi.nlm.nih.gov/pubmed/27563093}

\bibitem{Miyamoto2017}
{Yoshiyuki Miyamoto, Hong Zhang, Xinlu Cheng}, A.~Rubio,
  \href{https://link.aps.org/doi/10.1103/PhysRevB.96.115451
  https://journals.aps.org/prb/accepted/16078Ya6J771ca52b38b30e3c3ab6c4bad93aa5cd}{{Modeling
  of laser-pulse induced water decomposition on two-dimensional materials by
  simulations based on time-dependent density functional theory}} (sep 2017).
\newblock \href {http://dx.doi.org/10.1103/PhysRevB.96.115451}
  {\path{doi:10.1103/PhysRevB.96.115451}}.
\newline\urlprefix\url{https://link.aps.org/doi/10.1103/PhysRevB.96.115451
  https://journals.aps.org/prb/accepted/16078Ya6J771ca52b38b30e3c3ab6c4bad93aa5cd}

\bibitem{Hubener2017}
H.~H{\"{u}}bener, M.~A. Sentef, U.~{De Giovannini}, A.~F. Kemper, A.~Rubio,
  \href{http://www.ncbi.nlm.nih.gov/pubmed/28094286
  http://www.pubmedcentral.nih.gov/articlerender.fcgi?artid=PMC5247574
  http://www.nature.com/doifinder/10.1038/ncomms13940}{{Creating stable
  Floquet–Weyl semimetals by laser-driving of 3D Dirac materials}}, Nature
  Communications 8 (2017) 13940.
\newblock \href {http://arxiv.org/abs/1604.03399} {\path{arXiv:1604.03399}},
  \href {http://dx.doi.org/10.1038/ncomms13940}
  {\path{doi:10.1038/ncomms13940}}.
\newline\urlprefix\url{http://www.ncbi.nlm.nih.gov/pubmed/28094286
  http://www.pubmedcentral.nih.gov/articlerender.fcgi?artid=PMC5247574
  http://www.nature.com/doifinder/10.1038/ncomms13940}

\bibitem{Andrade2015}
X.~Andrade, D.~Strubbe, U.~{De Giovannini}, A.~H. Larsen, M.~J.~T. Oliveira,
  J.~Alberdi-Rodriguez, A.~Varas, I.~Theophilou, N.~Helbig, M.~J. Verstraete,
  L.~Stella, F.~Nogueira, A.~Aspuru-Guzik, A.~Castro, M.~A.~L. Marques,
  A.~Rubio, \href{http://xlink.rsc.org/?DOI=C5CP00351B
  http://pubs.rsc.org/en/content/articlehtml/2015/cp/c5cp00351b}{{Real-space
  grids and the Octopus code as tools for the development of new simulation
  approaches for electronic systems.}}, Physical chemistry chemical physics :
  PCCP 17~(47) (2015) 31371--31396.
\newblock \href {http://dx.doi.org/10.1039/c5cp00351b}
  {\path{doi:10.1039/c5cp00351b}}.
\newline\urlprefix\url{http://xlink.rsc.org/?DOI=C5CP00351B
  http://pubs.rsc.org/en/content/articlehtml/2015/cp/c5cp00351b}

\bibitem{A.Sato2014}
S.~A. Sato, K.~Yabana,
  \href{http://jlc.jst.go.jp/DN/JST.JSTAGE/jasse/1.98?lang=en{\&}from=CrossRef{\&}type=abstract
  https://www.jstage.jst.go.jp/article/jasse/1/1/1{\_}98/{\_}pdf}{{Maxwell +
  TDDFT multi-scale simulation for laser-matter interactions}} 1~(1) (2014)
  98--110.
\newblock \href {http://dx.doi.org/10.15748/jasse.1.98}
  {\path{doi:10.15748/jasse.1.98}}.
\newline\urlprefix\url{http://jlc.jst.go.jp/DN/JST.JSTAGE/jasse/1.98?lang=en{\&}from=CrossRef{\&}type=abstract
  https://www.jstage.jst.go.jp/article/jasse/1/1/1{\_}98/{\_}pdf}

\bibitem{Dewhurst2004}
J.~K. et.~al Dewhurst,
  \href{http://elk.sourceforge.net/}{http://elk.sourceforge.net} (2004).
\newline\urlprefix\url{http://elk.sourceforge.net/}

\bibitem{Groot2008}
F.~de~Groot, A.~Kotani, {Core level spectroscopy of solids}, 2008.

\bibitem{Gygi1995}
F.~Gygi, G.~Galli,
  \href{https://link.aps.org/doi/10.1103/PhysRevB.52.R2229}{{Real-space
  adaptive-coordinate electronic-structure calculations}}, Physical Review B
  52~(4) (1995) R2229--R2232.
\newblock \href {http://dx.doi.org/10.1103/PhysRevB.52.R2229}
  {\path{doi:10.1103/PhysRevB.52.R2229}}.
\newline\urlprefix\url{https://link.aps.org/doi/10.1103/PhysRevB.52.R2229}

\bibitem{Martin2008}
R.~M. Martin,
  \href{https://books.google.com/books/about/Electronic{\_}Structure.html?id=Sji3oQEACAAJ
  http://www.sudoc.fr/134215222}{{Electronic structure: basic theory and
  practical methods}}, Cambridge University Press, 2010.
\newline\urlprefix\url{https://books.google.com/books/about/Electronic{\_}Structure.html?id=Sji3oQEACAAJ
  http://www.sudoc.fr/134215222}

\bibitem{Schultz}
T.~Schultz, M.~Vrakking, {Attosecond and XUV Physics: Ultrafast Dynamics and
  Spectroscopy}, 2014.
\newblock \href {http://dx.doi.org/10.1002/9783527677689}
  {\path{doi:10.1002/9783527677689}}.

\bibitem{Emma2010}
P.~Emma, R.~Akre, J.~Arthur, R.~Bionta, C.~Bostedt, J.~Bozek, A.~Brachmann,
  P.~Bucksbaum, R.~Coffee, F.-J. Decker, Y.~Ding, D.~Dowell, S.~Edstrom,
  A.~Fisher, J.~Frisch, S.~Gilevich, J.~Hastings, G.~Hays, P.~Hering, Z.~Huang,
  R.~Iverson, H.~Loos, M.~Messerschmidt, A.~Miahnahri, S.~Moeller, H.-D. Nuhn,
  G.~Pile, D.~Ratner, J.~Rzepiela, D.~Schultz, T.~Smith, P.~Stefan,
  H.~Tompkins, J.~Turner, J.~Welch, W.~White, J.~Wu, G.~Yocky, J.~Galayda,
  \href{http://dx.doi.org/10.1038/nphoton.2010.176}{{First lasing and operation
  of an {\aa}ngstrom-wavelength free-electron laser}}, Nature Photonics 4~(9)
  (2010) 641--647.
\newblock \href {http://dx.doi.org/10.1038/nphoton.2010.176}
  {\path{doi:10.1038/nphoton.2010.176}}.
\newline\urlprefix\url{http://dx.doi.org/10.1038/nphoton.2010.176}

\bibitem{Goulielmakis2010}
E.~Goulielmakis, Z.-H. Loh, A.~Wirth, R.~Santra, N.~Rohringer, V.~S. Yakovlev,
  S.~Zherebtsov, T.~Pfeifer, A.~M. Azzeer, M.~F. Kling, S.~R. Leone, F.~Krausz,
  {Real-time observation of valence electron motion}, Nature 466~(7307) (2010)
  739--743.
\newblock \href {http://dx.doi.org/10.1038/nature09212}
  {\path{doi:10.1038/nature09212}}.

\bibitem{Leone2014}
S.~R. Leone, C.~W. McCurdy, J.~Burgd{\"{o}}rfer, L.~S. Cederbaum, Z.~Chang,
  N.~Dudovich, J.~Feist, C.~H. Greene, M.~Ivanov, R.~Kienberger, U.~Keller,
  M.~F. Kling, Z.-H. Loh, T.~Pfeifer, A.~N. Pfeiffer, R.~Santra, K.~Schafer,
  A.~Stolow, U.~Thumm, M.~J.~J. Vrakking,
  \href{http://www.nature.com/doifinder/10.1038/nphoton.2014.48}{{What will it
  take to observe processes in 'real time'?}}, Nature Photonics 8~(3) (2014)
  162--166.
\newblock \href {http://dx.doi.org/10.1038/nphoton.2014.48}
  {\path{doi:10.1038/nphoton.2014.48}}.
\newline\urlprefix\url{http://www.nature.com/doifinder/10.1038/nphoton.2014.48}

\bibitem{Schultze2014}
M.~Schultze, K.~Ramasesha, C.~D. Pemmaraju, S.~A. Sato, D.~Whitmore,
  A.~Gandman, J.~S. Prell, L.~J. Borja, D.~Prendergast, K.~Yabana, D.~M.
  Neumark, S.~R. Leone,
  \href{http://www.sciencemag.org/content/346/6215/1348}{{Attosecond band-gap
  dynamics in silicon}}, Science 346~(6215) (2014) 1348--1352.
\newblock \href {http://dx.doi.org/10.1126/science.1260311}
  {\path{doi:10.1126/science.1260311}}.
\newline\urlprefix\url{http://www.sciencemag.org/content/346/6215/1348}

\bibitem{Zurch2017}
M.~Z{\"{u}}rch, H.~T. Chang, P.~M. Kraus, S.~K. Cushing, L.~J. Borja,
  A.~Gandman, C.~J. Kaplan, M.~H. Oh, J.~S. Prell, D.~Prendergast, C.~D.
  Pemmaraju, D.~M. Neumark, S.~R. Leone,
  \href{http://aip.scitation.org/doi/10.1063/1.4985056}{{Ultrafast carrier
  thermalization and trapping in silicon-germanium alloy probed by extreme
  ultraviolet transient absorption spectroscopy}}, Structural Dynamics 4~(4)
  (2017) 044029.
\newblock \href {http://dx.doi.org/10.1063/1.4985056}
  {\path{doi:10.1063/1.4985056}}.
\newline\urlprefix\url{http://aip.scitation.org/doi/10.1063/1.4985056}

\bibitem{Moulet2017}
A.~Moulet, J.~B. Bertrand, T.~Klostermann, A.~Guggenmos, N.~Karpowicz,
  E.~Goulielmakis, \href{http://www.ncbi.nlm.nih.gov/pubmed/28912241
  http://www.sciencemag.org/lookup/doi/10.1126/science.aan4737}{{Soft x-ray
  excitonics}}, Science 357~(6356) (2017) 1134--1138.
\newblock \href {http://dx.doi.org/10.1126/science.aan4737}
  {\path{doi:10.1126/science.aan4737}}.
\newline\urlprefix\url{http://www.ncbi.nlm.nih.gov/pubmed/28912241
  http://www.sciencemag.org/lookup/doi/10.1126/science.aan4737}

\bibitem{Zhang2015}
Y.~Zhang, W.~Hua, K.~Bennett, S.~Mukamel,
  \href{http://link.springer.com/10.1007/128{\_}2014{\_}618}{{Nonlinear
  spectroscopy of core and valence excitations using short X-ray pulses:
  Simulation challenges}}, in: Density-Functional Methods for Excited States,
  Springer, Cham, 2015, pp. 273--345.
\newblock \href {http://dx.doi.org/10.1007/128_2015_618}
  {\path{doi:10.1007/128_2015_618}}.
\newline\urlprefix\url{http://link.springer.com/10.1007/128{\_}2014{\_}618}

\bibitem{Lopata2012}
K.~Lopata, B.~E. {Van Kuiken}, M.~Khalil, N.~Govind,
  \href{http://pubs.acs.org/doi/abs/10.1021/ct3005613}{{Linear-Response and
  Real-Time Time-Dependent Density Functional Theory Studies of Core-Level
  Near-Edge X-Ray Absorption}}, Journal of Chemical Theory and Computation
  8~(9) (2012) 3284--3292.
\newblock \href {http://dx.doi.org/10.1021/ct3005613}
  {\path{doi:10.1021/ct3005613}}.
\newline\urlprefix\url{http://pubs.acs.org/doi/abs/10.1021/ct3005613}

\bibitem{Soler2002}
J.~M. Soler, E.~Artacho, J.~D. Gale, A.~Garcia, J.~Junquera, P.~Ordejon,
  D.~Sanchez-Portal,
  \href{http://arxiv.org/abs/cond-mat/0111138{\%}0Ahttp://dx.doi.org/10.1088/0953-8984/14/11/302}{{The
  SIESTA method for ab initio order-N materials simulation}} 2745.
\newblock \href {http://arxiv.org/abs/0111138} {\path{arXiv:0111138}}, \href
  {http://dx.doi.org/10.1088/0953-8984/14/11/302}
  {\path{doi:10.1088/0953-8984/14/11/302}}.
\newline\urlprefix\url{http://arxiv.org/abs/cond-mat/0111138{\%}0Ahttp://dx.doi.org/10.1088/0953-8984/14/11/302}

\bibitem{Hohenberg1964}
P.~Hohenberg,
  \href{http://link.aps.org/doi/10.1103/PhysRev.136.B864}{{Inhomogeneous
  Electron Gas}}, Physical Review 136~(3B) (1964) B864--B871.
\newblock \href {http://dx.doi.org/10.1103/PhysRev.136.B864}
  {\path{doi:10.1103/PhysRev.136.B864}}.
\newline\urlprefix\url{http://link.aps.org/doi/10.1103/PhysRev.136.B864}

\bibitem{Kohn1965}
W.~Kohn, L.~J. Sham,
  \href{http://link.aps.org/doi/10.1103/PhysRev.140.A1133}{{Self-Consistent
  Equations Including Exchange and Correlation Effects}}, Physical Review
  140~(4A) (1965) A1133--A1138.
\newblock \href {http://dx.doi.org/10.1103/PhysRev.140.A1133}
  {\path{doi:10.1103/PhysRev.140.A1133}}.
\newline\urlprefix\url{http://link.aps.org/doi/10.1103/PhysRev.140.A1133}

\bibitem{Lebedev1999}
V.~I. D. N.~L. Lebedev, {A quadrature formula for the sphere of the 131st
  algebraic order of accuracy}, Doklady Mathematics. 53~(3) (1999) 477--481.

\bibitem{Crank1947}
J.~Crank, P.~Nicolson, D.~R. Hartree,
  \href{http://www.journals.cambridge.org/abstract{\_}S0305004100023197}{{A
  practical method for numerical evaluation of solutions of partial
  differential equations of the heat-conduction type}}, Mathematical
  Proceedings of the Cambridge Philosophical Society 43~(01) (1947) 50.
\newblock \href {http://dx.doi.org/10.1017/S0305004100023197}
  {\path{doi:10.1017/S0305004100023197}}.
\newline\urlprefix\url{http://www.journals.cambridge.org/abstract{\_}S0305004100023197}

\bibitem{salmon}
K.~Yabana, S.~A. Sato, M.~Noda, G.~F. Bertsch~et al, http://salmon-tddft.jp
  (2017).

\bibitem{Perdew1981}
J.~P. Perdew,
  \href{http://link.aps.org/doi/10.1103/PhysRevB.23.5048}{{Self-interaction
  correction to density-functional approximations for many-electron systems}},
  Physical Review B 23~(10) (1981) 5048--5079.
\newblock \href {http://dx.doi.org/10.1103/PhysRevB.23.5048}
  {\path{doi:10.1103/PhysRevB.23.5048}}.
\newline\urlprefix\url{http://link.aps.org/doi/10.1103/PhysRevB.23.5048}

\bibitem{Otobe2012}
T.~Otobe,
  \href{http://aip.scitation.org/doi/10.1063/1.4716192}{{First-principle
  description for the high-harmonic generation in a diamond by intense short
  laser pulse}}, Journal of Applied Physics 111~(9) (2012) 093112.
\newblock \href {http://dx.doi.org/10.1063/1.4716192}
  {\path{doi:10.1063/1.4716192}}.
\newline\urlprefix\url{http://aip.scitation.org/doi/10.1063/1.4716192}

\bibitem{wyckoff1963crystal}
R.~W.~G. Wyckoff,
  \href{https://books.google.com/books?id=40uGpwAACAAJ}{{Crystal Structures}},
  no. v. 1 in Crystal Structures, Wiley, 1963.
\newline\urlprefix\url{https://books.google.com/books?id=40uGpwAACAAJ}

\bibitem{Lee2012}
A.~J. Lee, F.~D. Vila, J.~J. Rehr,
  \href{https://link.aps.org/doi/10.1103/PhysRevB.86.115107}{{Local
  time-correlation approach for calculations of x-ray spectra}}, Physical
  Review B - Condensed Matter and Materials Physics 86~(11) (2012) 115107.
\newblock \href {http://dx.doi.org/10.1103/PhysRevB.86.115107}
  {\path{doi:10.1103/PhysRevB.86.115107}}.
\newline\urlprefix\url{https://link.aps.org/doi/10.1103/PhysRevB.86.115107}

\bibitem{Sato2015a}
S.~A. Sato, Y.~Taniguchi, Y.~Shinohara, K.~Yabana,
  \href{http://aip.scitation.org/doi/10.1063/1.4937379}{{Nonlinear electronic
  excitations in crystalline solids using meta-generalized gradient
  approximation and hybrid functional in time-dependent density functional
  theory}}, Journal of Chemical Physics 143~(22) (2015) 224116.
\newblock \href {http://arxiv.org/abs/1507.05156} {\path{arXiv:1507.05156}},
  \href {http://dx.doi.org/10.1063/1.4937379} {\path{doi:10.1063/1.4937379}}.
\newline\urlprefix\url{http://aip.scitation.org/doi/10.1063/1.4937379}

\bibitem{Burke2005}
K.~Burke, R.~Car, R.~Gebauer,
  \href{https://link.aps.org/doi/10.1103/PhysRevLett.94.146803}{{Density
  Functional Theory of the Electrical Conductivity of Molecular Devices}},
  Physical Review Letters 94~(14) (2005) 146803.
\newblock \href {http://dx.doi.org/10.1103/PhysRevLett.94.146803}
  {\path{doi:10.1103/PhysRevLett.94.146803}}.
\newline\urlprefix\url{https://link.aps.org/doi/10.1103/PhysRevLett.94.146803}

\end{thebibliography}

\end{document}